\documentclass[twocolumn,showpacs,preprintnumbers,amsmath,amssymb,superscriptaddress]{revtex4-2}

\usepackage{epsf}
\usepackage{graphicx}
\usepackage{sidecap}
 \usepackage{soul}
 \usepackage{array}
 \usepackage{amsmath}
 \usepackage{amssymb}
 \usepackage{dcolumn}
 \usepackage{epstopdf}
 \usepackage{bm}
 \usepackage{amsmath}
 
\usepackage{gensymb}
\usepackage{color}
\usepackage{hyperref}
\usepackage{soul}
\sethlcolor{green}
\usepackage{bbding}
\hypersetup{
    colorlinks=true,
    citecolor=red,
    linkcolor=red,
    filecolor=blue,   
    urlcolor=blue,
}
 
\def \beq {\begin{equation}}
\def \eeq {\end{equation}}
\renewcommand{\figurename}{\textbf{Fig.}}
\renewcommand{\thefigure}{{\textbf{\arabic{figure}}}}

\def\bibsection{\refname}
\renewcommand{\refname}{\noindent\textbf{References}}

\begin{document}
\title{Observation of Fermi arcs and Weyl nodes in a noncentrosymmetric magnetic Weyl semimetal}
 
\author{Anup Pradhan Sakhya} \affiliation{Department of Physics, University of Central Florida, Orlando, Florida 32816, USA}  
\author{Cheng-Yi Huang} \affiliation{Department of Physics, Northeastern University, Boston, Massachusetts 02115, USA}
\author{Gyanendra~Dhakal}\affiliation{Department of Physics, University of Central Florida, Orlando, Florida 32816, USA}  
\author{Xue-Jian Gao}\affiliation{Department of Physics, Hong Kong University of Science and Technology, Hong Kong, China}
\author{Sabin~Regmi} \affiliation{Department of Physics, University of Central Florida, Orlando, Florida 32816, USA}  
\author{Baokai Wang} \affiliation{Department of Physics, Northeastern University, Boston, Massachusetts 02115, USA}
\author{Wei Wen} \affiliation{Key Laboratory for Quantum Materials of Zhejiang Province, Department of Physics, Westlake University, Hangzhou, Zhejiang 310024, China}
\author{R. -H. He} \affiliation{Key Laboratory for Quantum Materials of Zhejiang Province, Department of Physics, Westlake University, Hangzhou, Zhejiang 310024, China}
\author{Xiaohan Yao} \affiliation{Department of Physics, Boston College, Chestnut Hill, Massachusetts 02467, USA}
\author{Robert Smith} \affiliation{Department of Physics, University of Central Florida, Orlando, Florida 32816, USA}  
\author{Milo Sprague}\affiliation{Department of Physics, University of Central Florida, Orlando, Florida 32816, USA}  
\author{Shunye Gao} \affiliation{Beijing National Laboratory for Condensed Matter Physics and Institute of Physics, Chinese Academy of Sciences, Beijing 100190, China}
\author{Bahadur~Singh} \affiliation{Department of Condensed Matter Physics and Materials Science, Tata Institute of Fundamental Research, Mumbai 400005, India}
\author{Hsin Lin} \affiliation{Institute of Physics, Academia Sinica, Taipei, Taiwan, Republic of China}
\author{Su-Yang Xu} \affiliation{Department of Chemistry and Chemical Biology, Harvard University, Cambridge, Massachusetts 02138, USA}
\author{Fazel Tafti} \affiliation{Department of Physics, Boston College, Chestnut Hill, Massachusetts 02467, USA}
\author{Arun~Bansil} \affiliation{Department of Physics, Northeastern University, Boston, Massachusetts 02115, USA}
\author{Madhab Neupane} \thanks{Corresponding author:\href{mailto:madhab.neupane@ucf.edu} {madhab.neupane@ucf.edu}}\affiliation{Department of Physics, University of Central Florida, Orlando, Florida 32816, USA}

\begin{abstract}
Weyl semimetal (WSM), a novel state of quantum matter, hosts Weyl fermions as emergent quasiparticles resulting from the breaking of either inversion or time-reversal symmetry. Magnetic WSMs that arise from broken time-reversal symmetry provide an exceptional platform to understand the interplay between magnetic order and Weyl physics, but only a few WSMs have been realized. Here, we identify CeAlSi as a new non-centrosymmetric magnetic WSM via angle-resolved photoemission spectroscopy (ARPES) and first-principles, density-functional theory based calculations. Our surface-sensitive vacuum ultraviolet ARPES data confirms the presence of surface Fermi arcs as the smoking gun evidence for the existence of the Weyl semimetallic state in CeAlSi. We also observe bulk Weyl cones at the Fermi arc terminations in CeAlSi using bulk-sensitive soft-X-ray ARPES measurements. These results implicate CeAlSi as   a unique platform for investigating exotic quantum phenomena resulting from the interaction of topology and magnetism.
{\noindent }
\end{abstract}

\maketitle 

\indent The recent discovery of Weyl fermions as exotic quasiparticles in Weyl semimetals (WSMs) has generated tremendous research interest in condensed- matter physics and materials science, due to their intimate link between the concepts of different fields of physics and their technological applications \cite{Wan, Burkov, Franchini, Bahadur, Weng, Chen, Xu, Ding, Lv, Vanderbilt}. In a WSM, the conduction and valence bands disperse linearly in three-dimensional (3D) momentum space, and the point where they intersect is known as the Weyl node \cite{Lv, Alidoust, Felser}. The essential prerequisite for a WSM is that either the crystal inversion (CI) symmetry \cite{Weng} or the time-reversal (TR) symmetry \cite{Wan} must be broken. The TaAs-family of transition-metal pnictides \cite{Weng, Xu} and the MoTe$_2$ and WTe$_2$ family of transition-metal dichalcogenides \cite{Soluyanov, Sun, Deng, Kaminski} belong to the class of WSMs with broken CI symmetry. TR symmetry broken WSMs were discovered recently in ferromagnets such as Co$_3$Sn$_2$S$_2$, Co$_2$MnGa, YbMnBi$_2$ \cite{Shen, Shekhar, Kim, Hasan, Borisenko}. However, when both symmetries are broken, a rare WSM arises that provides an exciting opportunity to understand the correlation between magnetism and Weyl fermions \cite{Chang, Singh}.\linebreak
   \indent Recently, RAlGe and RAlSi (R = Rare earth elements) materials, which are noncentrosymmetric, have been proposed as a new family of magnetic WSMs, where the Weyl nodes are tunable due to intrinsic magnetic order \cite{LaAlGe, CeAlGe, LaAlSi, CeAlGer, NdAlSi, CeAlSi, PrAlGe}. These families of compounds can exhibit different types of magnetic ground states and host Type-I or Type-II Weyl fermions. CeAlSi, a material from the RAlSi family, was recently shown to be a noncentrosymmetric ferromagnetic WSM that exhibits an anisotropic anomalous Hall effect between the easy and hard magnetic axes \cite{Singh}. The large ferromagnetic domains observed below T$_c$ = 8.3 K, make this material exciting as the Weyl nodes can be generated and manipulated by breaking both the CI and TR symmetries, making it an ideal material to explore the interplay between Weyl electrons and magnetism.

\begin{figure*} 
	\includegraphics[width=16cm]{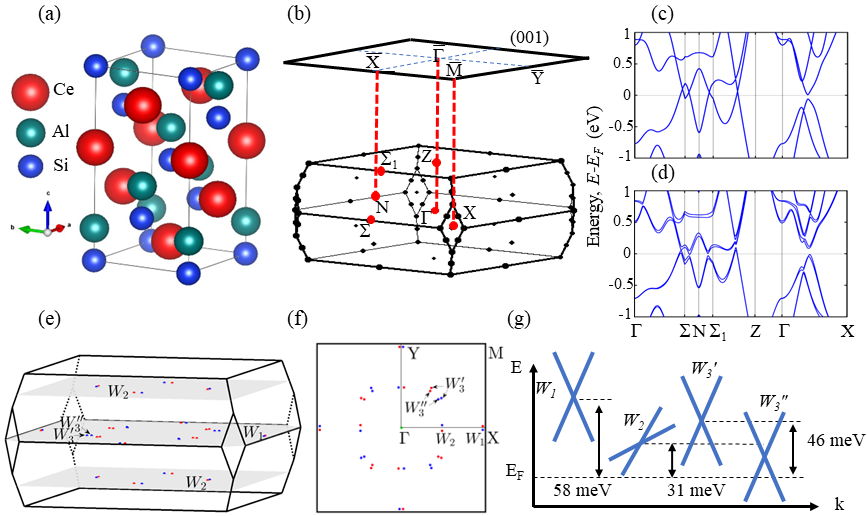} 
    \vspace{-1ex}
	\caption{Crystal structure and electronic structure of CeAlSi. (a) Unit cell of CeAlSi where the red, teal, and blue colored solid spheres denote the Ce, Al and Si atoms respectively. (b) Bulk Brillouin zone (BZ) and the projected (001) surface BZ. (c) Calculated bulk bands along the high symmetry directions without SOC and (d) with the inclusion of SOC. (e) Configuration of the 40 Weyl nodes in the bulk BZ created upon the inclusion of SOC. (f) Projection of the Weyl nodes on the (001) surface BZ in one quadrant highlighting all the Weyl nodes. The red (blue) dots denote the Weyl node with chirality +1 (-1) respectively. (g) Schematics comparing the three types of Weyl nodes appearing upon the inclusion of SOC.}
\label{fig1}
\end{figure*}
\indent In this Letter, by performing vacuum ultraviolet (VUV) and soft X-ray (SX) angle-resolved photoemission spectroscopy (ARPES) measurements, we report a systematic study of the electronic structure of CeAlSi in the paramagnetic phase. The surface-sensitive VUV ARPES results confirm the presence of  Fermi arcs, signatures of the WSMs on the (001) surface of CeAlSi. Bulk sensitive SX ARPES measurements reveal the presence of Weyl nodes in the paramagnetic phase. The experimental results are well corroborated by our density functional theory (DFT) results. We thus establish CeAlSi as a non-centrosymmetric topological WSM material, a platform for investigating the interplay between Weyl phases and magnetism.

\begin{figure*} 
	\centering
	\includegraphics[width=17cm]{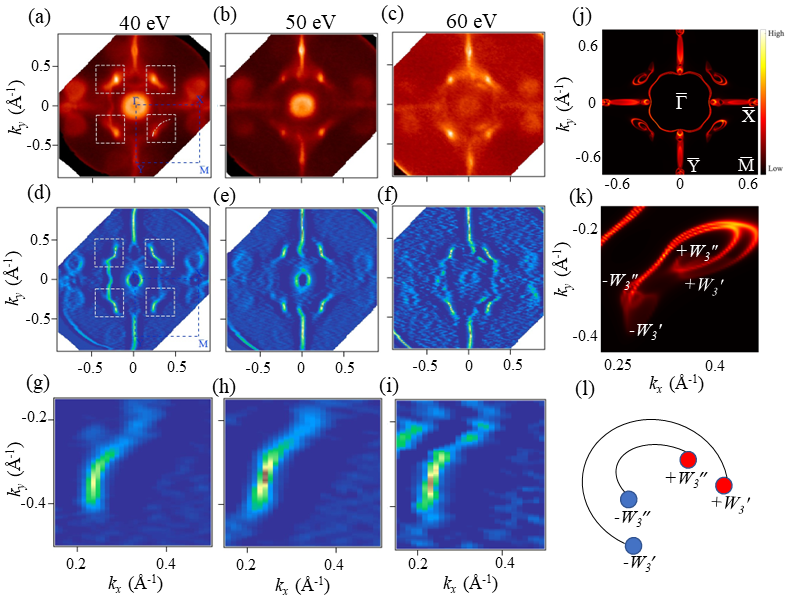}
   \vspace{-1ex}

\caption {Fermi surface as a function of photon energy. Experimental Fermi surface measured along the (001) direction using a photon energy of (a) 40 eV, (b) 50 eV, and (c) 60 eV. The spectral intensity is integrated within 20 meV of the chemical potential. Blue dashed lines represent the constructed Brillouin zone in one quadrant. The squares (white dashed lines) along the $\overline{\Gamma}–\overline{M}$ direction indicate the Fermi arcs in CeAlSi. The white dashed curves serves as guides to the eye to visualize the Fermi arcs in one quadrant of the Brillouin zone. (d-f) Second-derivative plots of the Fermi surface as shown in (a-c) respectively. The white dashed squares along the $\overline{\Gamma}–\overline{M}$ direction highlights the Fermi arcs. (g-i) Zoomed view of the ARPES measured Fermi arcs in one quadrant of the Brillouin zone. (j) Calculated Fermi surface along the (001) direction. (k) Zoomed view of the calculated Fermi surface highlighting the Fermi arcs. (l) Schematic illustration showing Fermi arcs (black lines) emerging from W3$\rq$ or W3$\rq$$\rq$ Weyl nodes along the $\overline{\Gamma}-\overline{M}$ direction. All measurements were performed at a temperature of 15 K.}
\label{fig2}
\end{figure*}

\begin{figure*}
\centering
\includegraphics[width=16cm]{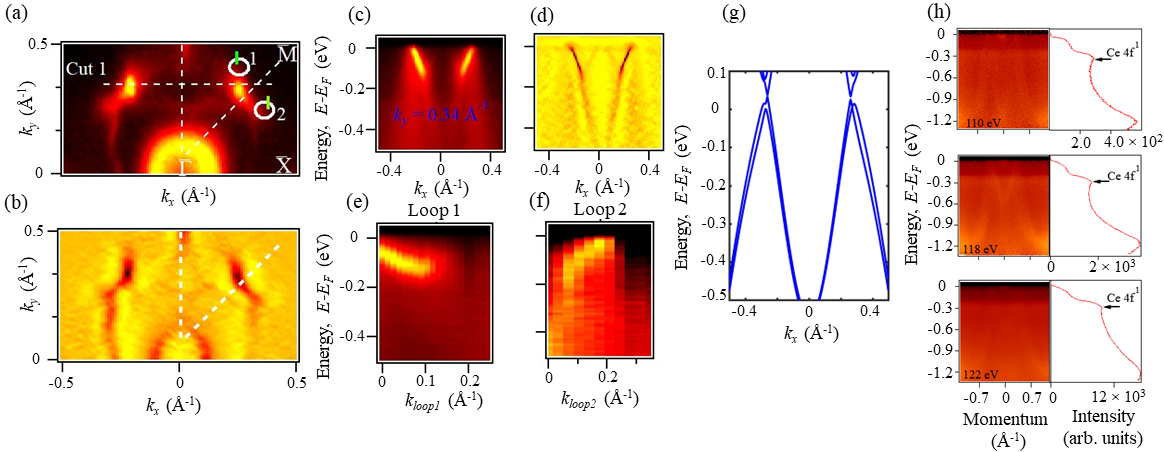} 
\vspace{-1ex}
\caption{Observation of Chiral charges in CeAlSi. Zoomed FS (a) and the curvature plot (b) obtained from ARPES measurement with a photon energy of 40 eV. Cuts of interest are illustrated with horizontal line labeled Cut 1 and white loops labeled 1 and 2 with the starting/end points marked by the vertical green lines. (c) Measured band dispersion along horizontal Cut 1. (d) Second derivative plot of (c). (e) ARPES measured band dispersion along the loop 1. Loop 1 encloses the termination point of the measured Fermi arc and shows a single left-moving chiral mode, corresponding to an enclosed Chern number n $ = $ -1 and (f) Band dispersion along the loop 2 showing a single right-moving chiral mode, corresponding to a Chern number n = +1. (g) Calculated energy dispersion around \textit{$k_y$} = 0.34 \AA $^{-1}$. (h) Band dispersion measured along the $\overline{\Gamma}-\overline{M}$ direction and its corresponding integrated energy distribution curves showing the presence of Ce 4\textit{f} states near the Fermi level at various photon energies as noted on the figure. All measurements were performed at a temperature of 15 K.}
\label{fig4}
\end{figure*}

\indent CeAlSi single crystals were grown by a self-flux method using Canfield crucible sets \cite{canfield}. For details of the single crystal growth technique and ARPES measurements, see the Supplemental Material \cite{SM}. Density functional theory (DFT) calculations were performed using the experimental lattice parameters (a = 4.25 Å; c = 14.58 Å) and the projector-augmented-wave (PAW) method implemented in the Vienna ab initio simulation package (VASP) \cite{VASP}. The exchange-correlation effects were included by using the generalized gradient approximation (GGA). The spin-orbit coupling (SOC) was included self-consistently \cite{Kresse, Perdew}. An on-site Coulomb interaction was added for Ce \textit{f}-electrons within the GGA + U scheme with U$_{eff}$ = 6 eV. A Wannier tight-binding Hamiltonian was obtained from the ab initio results using the VASP2WANNIER90 interface, which was subsequently used in our topological properties calculations \cite{Marzari}. CeAlSi is a semimetallic material that crystallizes in the tetragonal structure with space group \textit{I}4$_1$\textit{md} (No. 109) with lattice constants a = 4.25 \AA, and c = 14.58 \AA~ \cite{Singh}, as illustrated in Fig. \ref{fig1}(a). The crystal structure has two vertical mirror planes as well as two vertical glide mirror planes, but it lacks a horizontal mirror plane, thus breaking the inversion symmetry. Figure \ref{fig1}(b) shows the three-dimensional bulk Brillouin zone (BZ) of CeAlSi, and the projected two-dimensional BZ onto the (001) surface. The calculated bulk band structures along the high symmetry directions without spin-orbit coupling (SOC) and with SOC are presented in Figs. \ref{fig1}(c) and Fig. \ref{fig1}(d), respectively. The Fermi level is indicated with a horizontal line at E = 0 eV. From the SOC-included calculations, we can clearly see a hole pocket near the $\Sigma$ high-symmetry point and along the $\Sigma_1$-Z direction. The Weyl nodes in the \textit{k$_z$} = 0 plane are pinned by the C$_2$T symmetry. When SOC is included, there are four pairs of W1, W3$\rq$ and W3$\rq$$\rq$, as well as eight pairs of W2 Weyl nodes, as required by the C$_{4v}$ point group of the paramagnetic CeAlSi. The nodal lines on the \textit{k$_x$} = 0, \textit{k$_y$} = 0 mirror-invariant plane are stabilized by the mirror symmetry m$_x$, m$_y$, as the touching conduction and valence bands have opposite eigenvalues under the mirror operation. The 8 Weyl nodes on the \textit{k$_z$} = 0 plane are referred to as W1, whereas the remaining 16 Weyl nodes away from this plane are referred to as W2. Furthermore, each W3 (spinless) Weyl node breaks into two W3$\rq$ and W3$\rq$$\rq$ (spinful) Weyl nodes with the same chirality. As a result, there are a total of 40 Weyl nodes in the paramagnetic phase and the distribution of Weyl nodes in the bulk BZ and the (001) surface are presented in Figs. \ref{fig1}(e) and 1(f), respectively \cite{Chang}. The presence of Weyl nodes at various energies with respect to E$_F$ is shown in Fig. \ref{fig1}(g).

\begin{figure*}
\includegraphics[width=14cm]{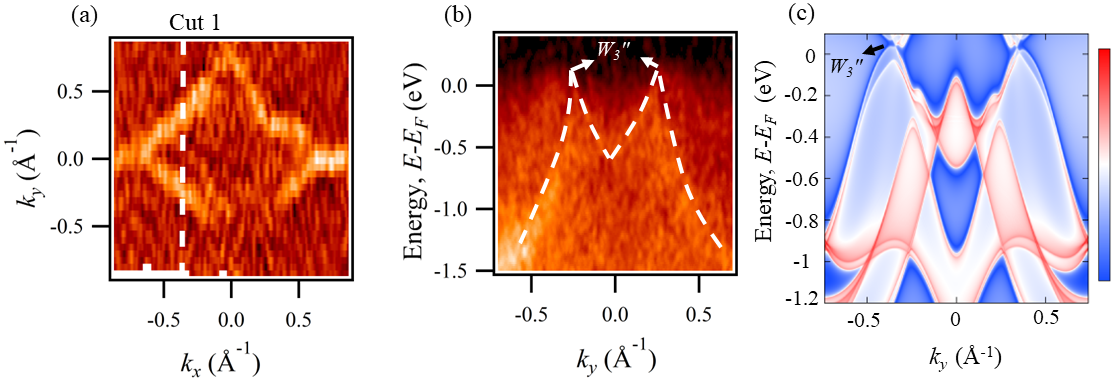}
\vspace{-1ex} 
\caption{Observation of bulk Weyl cone in CeAlSi. (a) Experimental FS on the \textit{k$_z$} = 0 plane obtained from SX-ARPES measurement with a photon energy of 498 eV.  (b) Experimental band dispersion measured along Cut 1 showing the linear dispersion of the W3$\rq$$\rq$ Weyl cone. The dashed line serves as guide to the eyes for better visualization. (c) DFT calculated bulk band projection around W3$\rq$$\rq$ along \textit{k$_y$}. The color marks the spectral intensity. All measurements were performed at a temperature of 20 K.}
\label{fig5}
 \end{figure*}
\indent To unveil the Weyl semimetallic phase in CeAlSi, we have presented the results from ARPES measurements on the (001) surface in Figs. 2-4. Fermi surface (FS) maps obtained from VUV-based ARPES measurements with different photon energies are presented in Figs. \ref{fig2}(a-c). The FS shows a circular pocket at the $\overline{\Gamma}$ point and mustache-like pockets at the $\overline{X}$ point and the $\overline{Y}$ points. These mustache-like pockets observed at the $\overline{Y}$ points are less intense compared to the pockets at the $\overline{X}$ points, and they can be seen distinctly at around 40 eV. To distinguish the surface and bulk contributions from various bands, we have performed photon-energy-dependent ARPES experiments using photon energy range from 40-60 eV. The measurements have been performed at a temperature of 15 K i.e., in the paramagnetic phase. The pockets observed at the $\overline{\Gamma}$, $\overline{X}$ and $\overline{Y}$ points disperse quite strongly with photon energy, suggesting their bulk 3D nature. Perhaps the most interesting observation is the presence of Fermi arcs along the $\overline{\Gamma}-\overline{M}$ direction, as can be seen in Figs. 2 (a-c). These Fermi arcs do not change shape and size with photon energy, suggesting their two-dimensional nature. This can be seen clearly in the photon-energy- dependent cuts taken along the $\overline{\Gamma}-\overline{M}$ direction where the Fermi arcs have been observed, and is presented in the Supplemental Material \cite{SM}.
Figures. 2(d)-2(f) shows the second-derivative plot of the FS map of Figs. 2(a)-2(c), respectively, which highlights the Fermi arcs along the $\overline{\Gamma}$-$\overline{M}$ direction. To better visualize the Fermi arcs, we have also presented the zoomed Fermi arcs in one quadrant of the Brillouin zone [see Figs. 2(g)-(i)] for photon energies of 40, 50, and 60 eV, respectively, where the Fermi arcs can be observed. The experimental FS is reasonably captured by our first-principles calculations, as can be seen in Fig. 2(j). Some discrepancies can be attributed to matrix element effects and the limitations of DFT in capturing the electronic structure of correlated electron systems \cite{NdSb, Gyan}. The zoomed view of the DFT obtained Fermi arc is presented in Fig. 2(k) for better visualization, and a schematic of the Fermi arc connecting the Weyl points as observed from DFT is presented in Fig. 2(l). 

\indent To prove that the arcs observed along the $\overline{\Gamma}-\overline{M}$ direction are indeed the Weyl Fermi arcs, we have examined the signature of projected chiral charge on the surface corresponding to a Chern number on the bulk, taking into account the bulk-boundary correspondence between the bulk Weyl fermions and surface Fermi arcs \cite{Wan, Belopolski}. For this we have considered closed loops in the surface BZ along one of the Fermi arcs, as shown in Fig. \ref{fig4}(a) by white circles designated as loop 1 and loop 2. The curvature plot of the Fermi surface shown in Fig. \ref{fig4}(a) is presented in Fig. \ref{fig4}(b) to observe the Fermi arcs distinctly. The band dispersion and its second derivative plots along the Cut 1 of Fig. \ref{fig4}(a), which lies on the \textit{$k_y$} = 0.34 \AA $^{-1}$ plane, are presented in Fig. \ref{fig4}(c) and Fig. \ref{fig4}(d), respectively. From Figs. \ref{fig4}(c) and \ref{fig4}(d) we can distinctly observe two counter propagating surface states. By analysing the energy-momentum dispersion for loop 1 (see, Fig. \ref{fig4}(a)), we detect one left-moving chiral mode which disperses towards E$_F$, thus showing chiral charge -1 on the associated bulk manifold, and it can be seen in Fig. \ref{fig4}(e). Similarly for loop 2, we  detect a right-moving chiral mode suggesting chiral charge of +1 as can be seen in Fig. \ref{fig4}(f). In this way, using only the surface sensitive ARPES measurements, we confirm that the arc-like features are indeed the Weyl Fermi arcs, and we measure the chiral charges in CeAlSi through the bulk-boundary correspondence \cite{Wan, Belopolski}. The calculated bulk band dispersion for a particular value of \textit{$k_y$} = 0.34 \AA $^{-1}$ is presented in Fig. \ref{fig4}(g) and is in agreement with the observed ARPES results. In Fig. \ref{fig4}(h), we have presented the ARPES measurements along the $\overline{\Gamma}-\overline{M}$ direction and its integrated energy distribution curves where we can clearly observe a sharp dispersionless peak at $\sim$ -0.3 eV below Fermi energy which corresponds to the so-called Kondo resonance peak and is labeled as 4\textit{f}$^{1}_{7/2}$ \cite{kondo}, suggesting that electronic correlation is important in the system. The 4\textit{f}$^{1}_{5/2}$ peak which lies near the Fermi level is not observed in the experiments. Further temperature-dependent and resonant photoemission experiments should be performed to understand the importance of Kondo physics in this system.\linebreak
\indent We have also performed bulk-sensitive SX-ARPES studies on CeAlSi to demonstrate the existence of the bulk Weyl cones and the Weyl nodes, as shown in Fig. \ref{fig5}. The SX-ARPES increases the relative spectral weight of bulk versus surface features because the photoelectrons excited by soft x-rays have a longer escape depth than the photoelectrons excited by VUV light. This SX-ARPES data can be readily compared with the corresponding bulk band calculations. In Fig. \ref{fig5}(a), we present the FS obtained from SX-ARPES measurements at 20 K on the \textit{k$_z$} = 0 plane with a photon energy of 498 eV. To observe the linear dispersion of the bulk Weyl cones, band energy dispersion along the cut 1 (indicated in the figure by dashed line) is shown in Fig. \ref{fig5}(b), that intersects the experimentally observed Fermi arcs and calculated position of the Weyl fermions. We observe linearly dispersing Weyl cones that correspond to the W3$\rq$$\rq$ Weyl node. Within our experimental resolution, the W3$\rq$$\rq$ Weyl fermion is located at (-0.32, -0.26)\AA $^{-1}$ on the \textit{k$_z$} = 0 plane. The bulk band dispersions shown in Fig. \ref{fig5}(b) along cut 1 are consistent with the DFT calculation presented in Fig. \ref{fig5}(c), which further confirms the presence of W3$\rq$$\rq$ Weyl cone along the cut 1 direction with the W3$\rq$$\rq$ Weyl node lying very close to E$_F$. Additional soft x-ray ARPES measurements performed at various photon energies further support our observation of the presence of W3$\rq$$\rq$ Weyl nodes in CeAlSi; see the Supplemental Material \cite{SM}. \linebreak
\indent In conclusion, our systematic study of the electronic band structure of CeAlSi using both VUV and SX ARPES reveals it as a CI symmetry-breaking Weyl semimetal. By utilizing VUV ARPES, on the (001) surface, we identified and confirmed the presence of the Fermi arcs, which is the conclusive evidence of the  Weyl semimetallic state. From SX-ARPES measurements, we were able to resolve linearly dispersing conical features that correspond to the lower part of the W3$\rq$$\rq$ Weyl cone. The experimental observations are well reproduced by our first-principles, DFT-based computations, further confirming the presence of Fermi arcs and Weyl cones in our experimental results. A recent report suggested that the breaking of TRS in this material will shift the position of the Weyl nodes in the BZ \cite{Singh}. The ARPES data which have been presented in this work is in the paramagnetic phase, i.e, the Weyl nodes are generated due to the broken inversion symmetry. CeAlSi is therefore a WSM with broken CI inversion symmetry. Having a magnetic transition at around 8.2 K, this system provides an excellent platform to study and understand the tuning of the Weyl points and the associated states with magnetism \cite{Singh}. The experimental confirmation for the shifting of the Weyl nodes due to the breaking of TRS is not yet there and remains a further point of our research interest. However, the present study confirms the Weyl Fermi arcs and the Weyl nodes in the paramagnetic phase arising from the breaking of inversion symmetry, and it will be useful for future studies as it can be used for comparison with those of the magnetic regime. Therefore, our study will open up a new avenue to study and unveil the theoretically suggested tuning of the Weyl nodes with the onset of magnetism, which is currently one of the most exciting topics in topological physics.\\

\noindent \textbf{ACKNOWLEDGMENTS}\\
\noindent M.N. acknowledges the  support from the Air Force Office of Scientific Research under Award No. FA9550-17-1-0415, the Air Force Office of Scientific Research MURI (Grant No. FA9550-20-1-0322), and the National Science Foundation (NSF) CAREER award DMR-1847962. The work at Northeastern University was supported by the Air Force Office of Scientific Research under award number FA9550-20-1-0322, and it benefited from the computational resources of Northeastern University's Advanced Scientific Computation Center (ASCC) and the Discovery Cluster. The work at Westlake University was supported by the National Natural Science Foundation of China (Grant No. 11874053) and Zhejiang Provincial Natural Science Foundation of China (LZ19A040001). This material is based upon work supported by the Air Force Office of Scientific Research under award number FA2386-21-1-4059. S.Y.X. is supported by NSF Career under award No. DMR-2143177. We thank D. Lu, M. Hashimoto, D. Hong and Z. H. Chen for beamline assistance. Use of the Stanford Synchrotron Radiation Lightsource, SLAC National Accelerator Laboratory, is supported by the U.S. Department of Energy, Office of Science, Office of Basic Energy Sciences under Contract No. DE-AC02-76SF00515. We thank Jessica L. McChesney for beamline assistance. This research used resources of the Advanced Photon Source, a U.S. Department of Energy (DOE) Office of Science User Facility at Argonne National Laboratory and is based on research supported by the U.S. DOE Office of Science-Basic Energy Sciences, under Contract No. DE-AC02-06CH11357.\\

\def\bibsection{\section*{\refname}}

\vspace{3ex}
\noindent \textbf{Competing interests}\\
The authors declare no competing interests.\\

\vspace{3ex}
\noindent \textbf{ADDITIONAL INFORMATION}\\

\textbf{Correspondence} and requests for materials should be addressed to Madhab Neupane.
\clearpage
\widetext
\begin{center}
\textbf{\large Supplemental Materials for \\~\\Observation of Fermi arcs and Weyl nodes in a noncentrosymmetric magnetic Weyl semimetal}
\end{center}
\setcounter{equation}{0}
\setcounter{figure}{0}
\setcounter{table}{0}
\setcounter{page}{1}
\makeatletter
\renewcommand{\theequation}{S\arabic{equation}}
\renewcommand{\thefigure}{S\arabic{figure}}
\renewcommand{\bibnumfmt}[1]{[#1]}
\renewcommand{\citenumfont}[1]{#1}
\renewcommand{\figurename}{{Supplementary Fig.}}
\renewcommand{\thefigure}{{{\arabic{figure}}}}
\renewcommand{\tablename}{Supplementary Table}
\renewcommand{\thetable}{\arabic{table}}
\def\bibsection{\refname}
\renewcommand{\refname}{\noindent\textbf{Supplementary References}}

\begin{center}\textbf{SINGLE CRYSTAL GROWTH}
\end{center}
\noindent High-quality single crystals of CeAlSi were grown by using self-flux method in Canfield crucible sets \cite{canfield} where the starting materials were weighed in the ratio Ce:Al:Si = 1:10:1. The mixture was placed inside a crucible in an evacuated quartz tube. The sealed tube was then heated to 1000 \degree C at the rate of 3 \degree C/min; this temperature was maintained for 12 h, and then cooled to 700 \degree C very slowly at 0.1 \degree C/min. The sealed tube was further kept at 700 \degree C for 12 h, and finally centrifuged to decant the residual Al flux.\\

\begin{center}\textbf{ANGLE-RESOLVED PHOTOEMISSION SPECTROSCOPY}
\end{center}
\noindent Angle-resolved photoemission spectroscopy (ARPES) measurements were performed at Stanford Synchrotron Radiation Lightsource (SSRL), beamline 5-2, Menlo park, CA. Measurements were carried out at a temperature of 15 K. The pressure in the UHV was maintained better than 1$\times$10$^{-10}$ torr. The angular and energy resolution were set better than 0.2\degree and 15 meV, respectively. Soft X-ray ARPES measurements were performed at the end station sector 29-ID in APS Beamline equipped with a hemispherical Scienta R4000 electron analyzer. The energy and angular resolutions were set better than 200 meV and 0.1{\degree}, respectively. The samples were cleaved at 20 K and the pressure was better than 1$\times$10$^{-10}$ torr. Additional Soft x-ray ARPES measurements were performed at the Dreamline of the Shanghai Synchrotron Radiation Facility (SSRF). The energy and angular resolutions were set better than 100 meV and 0.2{\degree}, respectively. The temperature was maintained at 20 K.\\

\begin{center}\textbf{CORE-LEVEL ANGLE-INTEGRATED PHOTOEMISSION SPECTROSCOPY}
\end{center}
Supplementary Fig. 1 shows the angle-integrated photoemission spectrum for CeAlSi over a wide energy range of binding energy fom the E$_F$ till 260 eV. The spectra shows the peaks of Ce, Al and Si core levels, confirming that our samples are composed of these three elements. 

\begin{figure*} [h!]
\centering
\includegraphics[width=12cm]{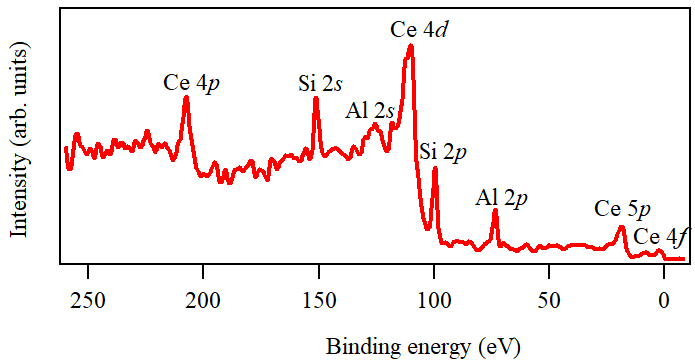}
\caption{Core level measurement of CeAlSi, which clearly shows the presence of Ce, Al, and Si peaks in the studied sample.} 
\end{figure*}

\begin{figure*} [h!]
\centering
\includegraphics[width=15cm]{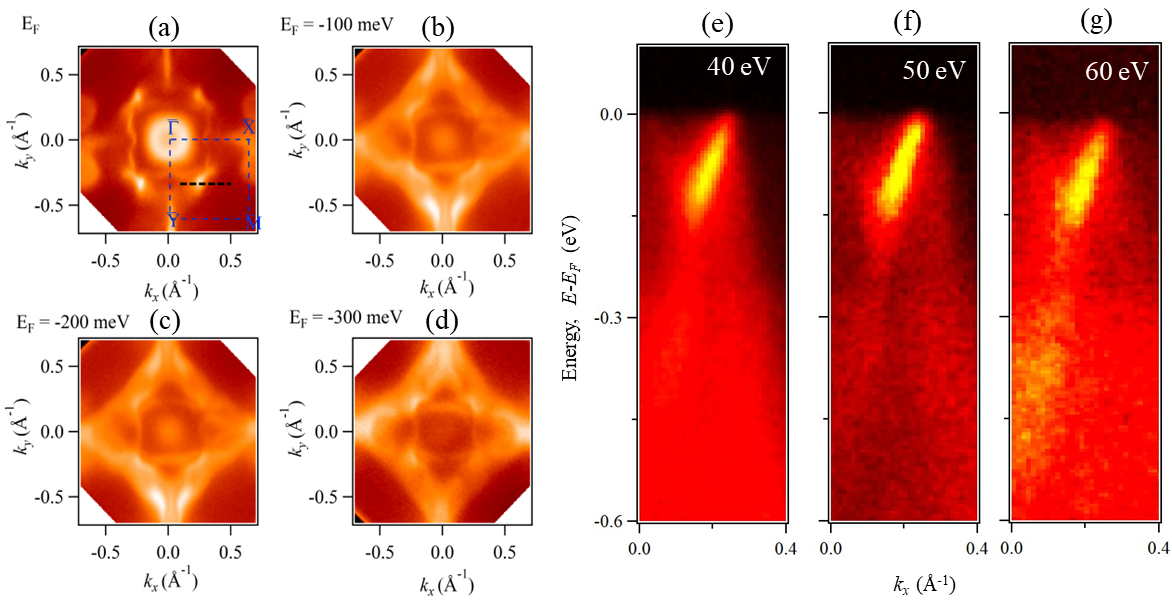} 
\caption{(a) Fermi surface and (b-d) constant energy contour plots at 15 K measured using a photon energy of 40 eV where the Fermi arcs can be seen along the $\overline{\Gamma}–\overline{M}$ direction. (e-g) Photon-energy-dependent ARPES along the horizontal line (black dotted colour in Supplementary Fig. 2a). Similar cuts were taken for Supplementary Fig. 2(f,g) (not shown). Negligible \textit{k$_z$} dispersion is observed for the Fermi arc. Data were taken at the SSRL beamline 5-2 at a temperature of 15 K.}
\label{fig3}
\end{figure*}

\begin{center}\textbf{CONSTANT ENERGY CONTOUR MAPS AND PHOTON ENERGY DEPENDENT MEASUREMENTS ALONG THE $\overline{\Gamma}-\overline{M}$ DIRECTION}
\end{center}
\indent In Supplementary Fig. 2(a-d), we present the Fermi surface (FS) map and the constant energy contour plots taken at various binding energies noted on top of each plot. The circular pocket at the $\overline{\Gamma}$ point contracts with increasing binding energy suggesting the  electron-like nature of the associated bands. The pockets at the $\overline{X}$ and $\overline{Y}$ points grow at higher binding energy indicating their hole-like nature. To better understand the nature of the Fermi arc feature along the $\overline{\Gamma}-\overline{M}$ direction, we have performed photon-energy-dependent studies. The band dispersion energy-momentum cut through the Fermi arcs (as shown by the black dotted line in Supplementary Fig. 2a for FS obtained at different photon energies) is presented in Supplementary Fig. 2(e-g) which suggests negligible dispersion with the change in photon energy from 40 eV to 60 eV, indicating its surface origin.

\begin{figure*} [h!]
\centering
\includegraphics[width=6cm]{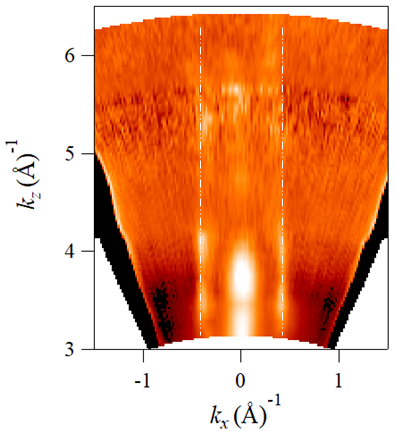}
\caption{\textit{k$_z$} dependent measurement along the $\overline{\Gamma}-\overline{M}$ high symmetry direction at the binding energy of 0 eV.}
\end{figure*}

\begin{center}\textbf{\textit{k$_z$} DEPENDENT MEASUREMENTS}
\end{center}
Supplementary Fig. 3 shows the photon-energy-dependent measurements along the $\overline{\Gamma}$-$\overline{M}$  high symmetry direction using photon energy ranging from 30 eV to 150 eV i.e., from {\textit{k$_{z}$} = 3 \AA$^{-1}$  to 6.5 \AA $^{-1}$. We have used V$_0$ = 12 eV to calculate the out of the plane momentum. The photon-energy-dependent measurements provide the evidence of the surface states of a material. The measurements have been performed at 15 K i.e, in the paramagnetic phase. The states present along the $\overline{\Gamma}$-$\overline{M}$ direction as indicated by the white-dashed lines does not disperse with photon energies thus confirming their two-dimensional nature and provide additonal support to our claim that these are the Fermi arcs in CeAlSi, whereas the states at the $\overline{\Gamma}$ show significant dispersion with photon energies suggesting their bulk nature. The photon energies used in the measurements provide extremely surface sensitive measurements. 

\begin{center}\textbf{ADDITIONAL SOFT X-RAY ARPES DATA}
\end{center}
We have performed additional Soft x-ray ARPES measurements on CeAlSi as shown in Supplementary Fig. 4. In Supplementary Fig. 4(a) we present the Fermi surface map measured using a photon energy of 500 eV where multiple Brillouin zones are covered in our experiments. The high  symmetry points are labeled in the figure. We have measuerd the Fermi maps in a narrow energy range as shown in Supplementary Fig. 4(b-l) using various photon energies. To confirm the presence of W3$\rq$$\rq$  Weyl cones, the band dispersion along the Cut 1 is presented in the lower panel of Supplementary Fig. 4(b-l). The  W3$\rq$$\rq$ linear Weyl cones with Weyl nodes lying very close to the E$_F$ can be clearly observed which supports our claim that CeAlSi is a Weyl semimetal in the paramagnetic phase. 

\begin{figure*} [h!]
\centering
\includegraphics[width=\textwidth]{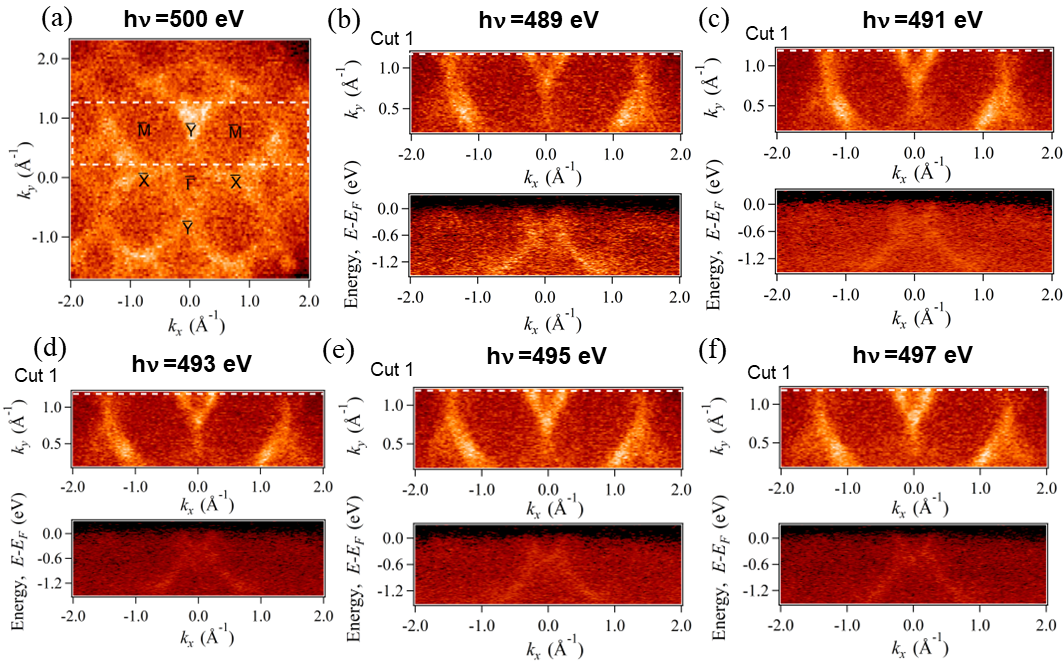}
\includegraphics[width=\textwidth]{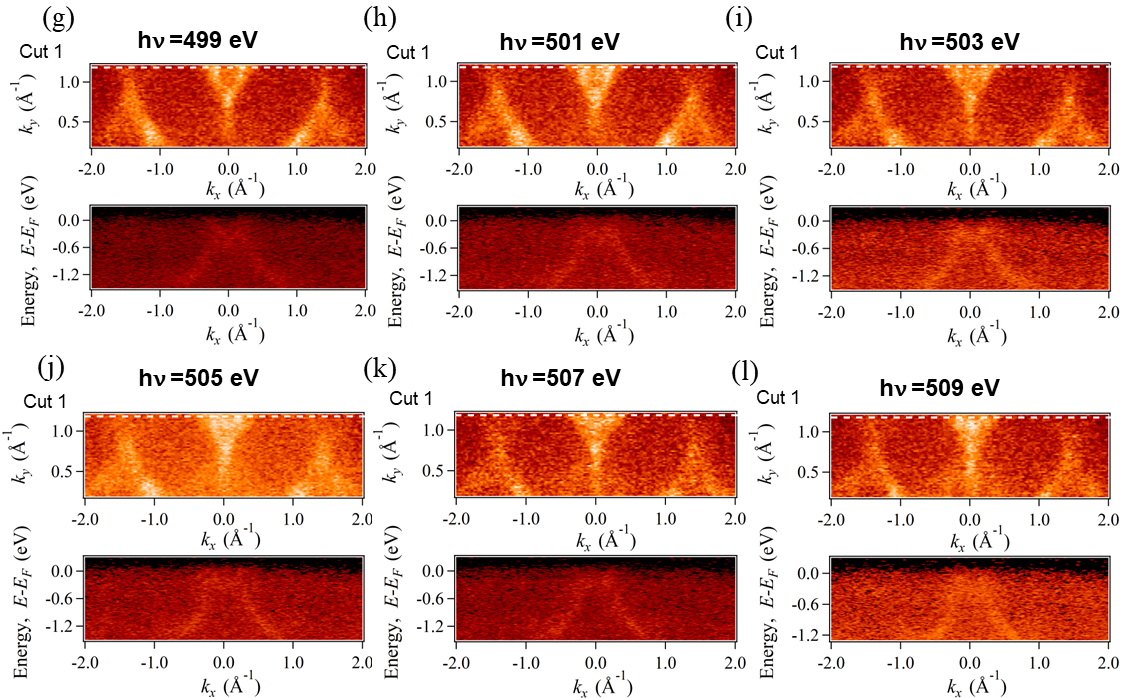}
\caption{Observation of bulk Weyl cones in CeAlSi at various photon energies. (a) Wide Fermi surface map measured using a photon energy of 500 eV where high-symmetry points of the Brillouin zone are marked. The rectangle (white dashed line) indicates the region of interest where the Fermi surface was measured for the data (b-l). (b) Soft X-ray Fermi surface measured using a photon energy of 489 eV (Upper panel) and Band dispersion along Cut 1 (white dashed lines) showing the linear dispersion of the Weyl cone (Lower panel). (c-l) same as b but measured at various photon energies as indicated in the respective figures.} 
\end{figure*}

\clearpage
\section*{Supplementary References:}
\justify
\bibitem{canfield} P. C. Canfield, T. Kong, U. S. Kaluarachchi, and N. H. Jo, \href{https://doi.org/10.1080/14786435.2015.1122248} {Philos. Mag. \textbf{96}, 84 (2016)}.

\end{document}